\documentclass[12pt]{iopart}
\usepackage{graphics}
\usepackage{epsfig}

\begin{document}

\title{Local scale invariance in the parity conserving nonequilibrium
  kinetic Ising model}
\author{G\'eza \'Odor}
\address{Research Institute for Technical Physics and Materials
  Science, H-1525 Budapest, P.O.Box 49, Hungary} 
\ead{odor@mfa.kfki.hu}

\begin{abstract}
Local scale invariance has been investigated in the 
nonequilibrium kinetic Ising model exhibiting absorbing phase
transition of PC type in 1+1 dimension. Numerical evidence has been
found for the satisfaction of this symmetry and estimates for the 
critical ageing exponents are given.
\end{abstract}
\maketitle

\section{Introduction}

The classification of the universality classes of non-equilibrium phase
transitions is still an open problem of statistical physics 
\cite{Dick-Mar,Hin2000,dok,kam}. In equilibrium conformal invariance (CI) 
\cite{FQS,CardyConf,HenkelConf} enables this in two dimensional critical
systems as the consequence of a larger group (the CI group) than the 
mere scale transformations. Recently the generalization 
of the generators of CI (albeit without invariance under time-translations)
are proposed for anisotropic, dynamical models \cite{LSIbev,HP05}. The 
corresponding invariance is the so-called local scale-invariance (LSI).
Since it is supposed to be the extension of the dynamical scale 
transformations for such systems it may serve as a convenient tool 
for classifying universality classes of non-equilibrium systems as well.

The quantities of main interest are the two-time autocorrelation function
$C(t,s)$ and the auto-response function $R(t,s)$, which describe ageing 
phenomena (for recent reviews see \cite{ageing}) 
\begin{eqnarray}
C(t,s) &=& \left\langle \phi(t,\vec{r}) \phi(s,\vec{r})\right\rangle \\
R(t,s) &=& \left. \frac{\delta\langle \phi(t,\vec{r})\rangle}{\delta
h(s,\vec{r})}  \right|_{h=0} 
\:=\: \left\langle  \phi(t,\vec{r}) \widetilde\phi(s,\vec{r})\right\rangle
\end{eqnarray}
where $\phi$ and $\widetilde\phi$ are the fields in the 
Janssen-de Dominicis formalism \cite{deDo78,Jans92} and $h$ is the 
magnetic field conjugate to $\phi$.
For $t,s\to\infty$ {\em and} $y=t/s>1$ one expects the scaling forms
\begin{eqnarray} \label{CRforms}
C(t,s) &=& s^{-b} f_C(t/s) \\
R(t,s) &=& s^{-1-a} f_R(t/s) ,
\end{eqnarray}
where $a$ and $b$ are ageing exponents and $f_C$ and $f_R$
are scaling functions such that $f_{C,R}(y)\sim y^{-\lambda_{C,R}/Z}$ for 
$y\gg 1$. Here $\lambda_C$ and $\lambda_R$ are the auto-correlation \cite{ace}
and auto-response \cite{are} exponents respectively and independent of 
equilibrium exponents and the dynamical exponent $Z$ 
(defined as usual $Z = \nu_{||} / \nu_{\perp}$).

As in case of CI one expects that LSI fully determines the 
functional form of the scaling functions. Henkel et al. derived
$R(t,s)$ in general and the form of $C(t,s)$ for $Z=2$
by identifying the quasi-primary operators of the theory \cite{HPGL,HEP0605}.
The generalized form of $R(t,s)$ takes into account the difference 
between physical observable defined in lattice models and the associated
quasi-primary scaling operators of the underlying field theory as well. 
This ansatz looks as
\begin{equation} \label{Rfullform}
R(t,s) = s^{-1-a} \left(\frac{t}{s}\right)^{1+a'-\lambda_R/Z}  
\left( \frac{t}{s}-1\right)^{-1-a'} \ ,
\end{equation}
where $a'\ne a$ is an independent ageing exponent in general.
Some systems with detailed balance symmetry has been analyzed 
recently and found to satisfy (\ref{Rfullform})  
\cite{HP05,Bert99,Maze04,Godr00,Maye06,Maye04,Plei05} with $a\ne a'$.
On the other hand renormalization-group results for some important
universality classes concluded that $a=a'$ should be hold.
In particular explicit two-loop field-theoretical computation 
of $R(t,s)$ for the $O(N)$ universality class and Model A dynamics 
at the critical point claim $a=a'$ \cite{CG02,CGJPA}.

Recently numerical simulations of the non-equilibrium contact process 
(CP) did not satisfy that form completely \cite{HinLSICP}
and Hinrichsen argued that LSI is not a generic property of ageing 
phenomena but is restricted to diffusive models ($Z=2$) or 
above the upper critical dimension.
In \cite{HEP0605} Henkel et al. suggested that there is crossover 
in case of nonequilibrium critical dynamics because both the ageing
regime ($t-s \sim O(s)$) and the quasi-stationary regime ($t-s<<s$)
display scaling with the same length scale $L(t) \propto t^{1/Z}$.
For a more detailed discussion of these results see a very recent
review \cite{HenkLSIrev}.

In this paper I present simulation results for an other
nonequilibrium critical model, the parity conserving (PC)
nonequilibrium Ising model (NEKIM) in $1+1$ dimensions. 
I provide numerical evidence that in this model $C(t,s)$ and $R(t,s)$ can be
fitted with the forms Eqs.(\ref{CRforms}),(\ref{Rfullform}) hence this
nonequilibrium critical model exhibits LSI scaling invariance.

\section{The PC class NEKIM model}

The NEKIM model has been introduced and analyzed first by 
Menyh\'ard \cite{Nora94} as a generalization of the Kinetic Ising model
\cite{Glauber} by adding spin-exchange updates in between the
spin-flipp steps of the Glauber Ising model.
In one dimension the domain walls (kinks) between up and down regions can
be considered as particles. The spin-flip dynamics can be mapped onto
particle movement
\begin{equation}
\uparrow\downarrow\downarrow\stackrel{w_i}{\rightleftharpoons}
\uparrow\uparrow\downarrow \ \
\sim \ \ \bullet\circ \rightleftharpoons \circ\bullet
\end{equation}
or to the annihilation of neighboring particles
\begin{equation}
\uparrow\downarrow\uparrow\stackrel{w_o}{\rightarrow}
\uparrow\uparrow\uparrow 
 \ \ \sim \ \ \circ\circ \rightarrow \bullet\bullet
\end{equation}
Therefore the $T=0$ Glauber dynamics is equivalent to the annihilating 
random walk (ARW). This is a double degenerate phase, an initial state decays 
algebraically to the stationary state, which is one of the absorbing
ones (all spins up or all spins down, provided the initial state 
has an even number of kinks).
By mapping the spin-exchange dynamics in the same 
way more complicated particle dynamics emerges, for example:
\begin{equation}
\uparrow\uparrow\downarrow\downarrow\stackrel{p_{ex}}{\rightleftharpoons}
\uparrow\downarrow\uparrow\downarrow \ \
\sim \ \ \circ\bullet\circ \rightleftharpoons \bullet\bullet\bullet
\end{equation}
one particle may give birth of two others or three particles may coagulate
to one. Therefore this model is equivalent to branching and annihilating
random walks with even number of offsprings \cite{Taka,Cardy-Tauber}.
By increasing the spin-exchange a second order phase transition takes 
place\cite{Nora94} for the {\it  kinks} from absorbing to active state, 
which belongs to the parity conserving (PC) universality class 
\cite{Gras84,Cardy-Tauber,Canet}.

In \cite{cpccikk} this model has been investigated by high precision
cluster simulations with the parameterization
\begin{eqnarray}
p_{ex}=1-2\Gamma \\
w_i=\Gamma (1-\delta)/2 \\ w_o=\Gamma (1+\delta) 
\end{eqnarray}
originating from the Glauber Ising model \cite{Glauber}.
In this paper I present simulations at the critical point 
determined in previous works \cite{Nora94,meorcikk,MeOd96,MeOd97,cpccikk}. 
The parameters chosen are:
$\Gamma=0.35$, $p_{ex}=0.3$, $\delta_c=-0.3928(2)$. Here
the kink ($n_i \in (0,1)$) density decays as 
\begin{equation} \label{densdec}
<n_i(t)> \propto t^{-0.285(2)} 
\end{equation}
as can be seen on the inset of Fig.\ref{LSIC}.
In previous works the NEKIM algorithm introduced as follows.
The spin-flip part was applied using two-sub-lattice updating. 
Following this states of the spins are stored and L (L is the size 
of the system) random attempts of spin-exchanges are done using the 
stored situation of states of the spins before updating the whole lattice.
All these together was counted as one Monte Carlo time-step (MCS)
of updating (throughout the paper time is measured by MCS).

\section{Simulations}

\begin{figure}
\begin{center}
\epsfxsize=130mm
\epsffile{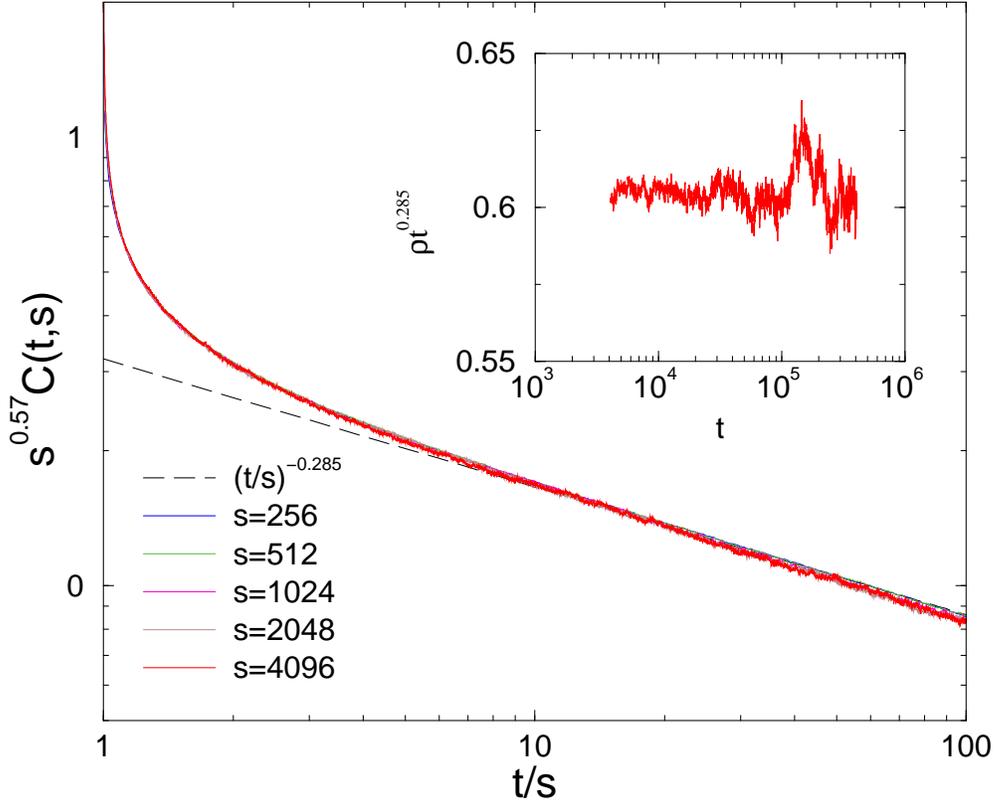}
\caption{Autocorrelation $C(t,s)$ of the critical
NEKIM model for several values of waiting time $s$ as a function
of the scaling variable $t/s$. The dashed line has the slope 
$-0.285\sim -\alpha$. The inset shows the kink density decay
$\rho(t)^{0.285}$ up to $t=409600$ MCS in a system of size
$L=10^5$.
\label{LSIC}}
\end{center}
\end{figure}

Time-dependent simulations were performed in $L=2\times 10^4 - 10^5$ 
sized systems with periodic boundary conditions. The runs were started
from fully ordered kink state ($n_i(0)=1$) i.e. alternating up-down spin
configuration ($s_i \in (-1,1)$). I followed the quench towards the 
critical state and measured the kink (order parameter) density
\begin{equation}
<n_i(t)> \propto t^{-\alpha} \ ,
\end{equation}
the kink-kink autocorrelation
\begin{equation}
C(t,s) = <n_i(t)n_i(s)> \ ,
\end{equation}
and the auto-response function, by flipping a spin at random site $l$ 
at time $s$ generating a kink pair out of the vacuum
\begin{equation}
R(t,s) = <n_i(t)n_{i+1}> - <n'_i(t)n'_{i+1}>  |_{s'_l(s) := -s_l(s)}
\end{equation}
The simulations were run for several values of waiting times 
$s=256, 512, 1024, 2048, 4096$ and the scaled autocorrelation 
$C(t,s) t^{-2\alpha}$ is plotted on Fig. \ref{LSIC} with 
the assumption of the form Eq.(\ref{CRforms}).
Good data collapse (within error margin of the simulations)
in case of systems of sizes $L=2\times 10^4$ could be achieved 
for the whole region, however for larger $s$ and $t$ values small 
deviations form the collapse could also be observed.
By investigating larger systems this proved to be finite size effect. 
The curve on Fig. \ref{LSIC} for $s=4096$ 
shows the result of $L=10^5$ simulations.
In the asymptotic $t/s\to\infty$ limit it can be fitted by
$t^{-0.285}$ power-law, corresponding to the density decay
of the PC class \cite{cpccikk}. This suggests the scaling exponents:
$b=0.570(4)$, $\lambda_C / Z = 0.285(2)$. By inserting the
value of the dynamical exponent of the PC class $Z=1.75(1)$
\cite{dok} on obtains $\lambda_C = 0.498(2)$.
This exponent agrees with that of the autocorrelation exponent
of spins $\lambda = 1.50(2)$ \cite{MeOd97} for $(t/s)\to\infty$
\begin{eqnarray}
A(t,s) & = & < s_i(s) s_i(t) > \nonumber \\
       & = & f(t/s) \propto (t/s)^{-(\lambda - d + 1 -\eta/2)/Z}  \ ,
\end{eqnarray}
since $\eta=1.01(1)$ and a hyperscaling-law connecting time
dependent spin and kink exponents at the PC transition point 
derived in \cite{MeOd96}. Fitting for the connected 
autocorrelation function defined as $\Gamma(t,s)=C(t,s)-N(t)N(s)$
resulted in $\lambda_G/Z=1.9(1)$

\begin{figure}
\begin{center}
\epsfxsize=130mm
\epsffile{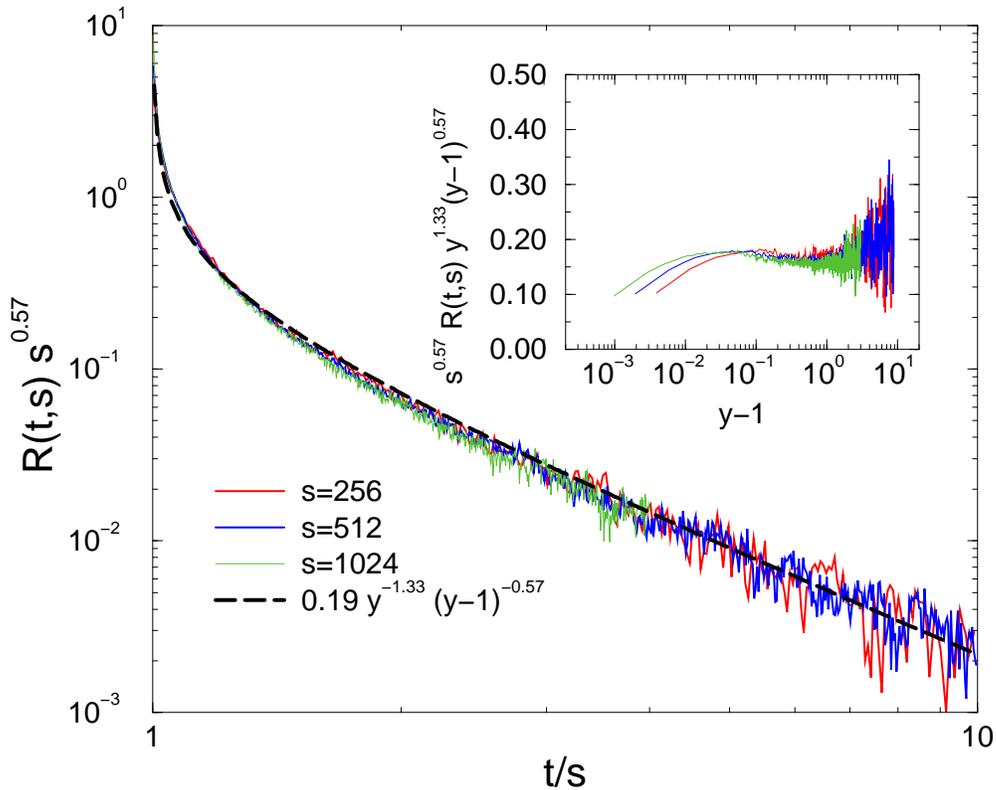}
\caption{Auto-response $R(t,s)$ of the critical
NEKIM model for several values of waiting time $s$ as a function
of the scaling variable $t/s$. The dashed line is a fit with the
general form (\ref{Rfullform}). In the inset the rescaled
auto-response function is plotted in such a way that possible 
deviations from the LSI scaling are more easily observable
for $y \to 1$. 
\label{LSIR}}
\end{center}
\end{figure}

The auto-response function has been found to exhibit similar nice data 
collapse by plotting  $R(t,s) t^{0.57}$ as the function of
$y = t/s$ (Fig.\ref{LSIR}). However in \cite{HinLSICP} Hinrichsen
discovered that in case of the CP model deviations from 
the LSI scaling form of $R(t,s)$ (\ref{Rfullform}) may occur for
$y\to 1$. To see this I plotted $R(t,s) s^{0.57} y^A (y-1)^B$
suggested in \cite{HinLSICP} as the function of $y-1$. 
Now one may see agreement with Eq.(\ref{Rfullform}) if the curves 
fitted with the parameters collapse and are horizontal 
for all $y$ values. The best agreement could be achieved 
with $A=-1.3(1)$ and $B=-0.57(1)$ plotted in the inset of
Fig.\ref{LSIR}. By increasing $s$ the observed deviations from
LSI scaling for $y \to 1$ occur at smaller $y$ values, 
suggesting corrections due to the microscopic reference time $s$. 
This is different from the case of the CP, where all such curves 
collapsed.
Assuming the general form Eq.(\ref{Rfullform}) fitting results 
in: $a = a' = -0.430(2)$, $\lambda_R/Z = 1.9(1)$ 
dynamical exponents with a validity of more than three decades.

\section{Conclusions}

In conclusion numerical simulations of the parity conserving
NEKIM model in 1d support local scale invariance at the critical point. 
In contrast to the contact process (belonging to the directed
percolation class \cite{Dick-Mar}) corrections to scaling due
to the microscopic reference time vanish in the $s\to\infty$ limit.
Both the autocorrelation and the auto-response functions 
can well be described by the functional forms of LSI, only
negligible dependence on the system sizes has been detected
within error margin of the numerical simulations. Further sources
of deviations may come from the value of $\alpha$ and the
location of the critical point. The same analysis done for
$-\delta_c = 0.3925, 0.392, 0.391$ and using $\alpha = 0.286, 0.287$ 
did not result in viewable deviations in the figures and the
fitting parameters.
The auto-response function scales in such a way that
$a = a'$. Numerical estimates for the $\lambda_{C,G,R}$
exponents are determined and $\lambda_G=\lambda_R=Z(1+\alpha)+d$
in agreement with the scaling hypothesis of \cite{CGK06}.
The results support the conjecture of Henkel \cite{HPGL} that 
LSI can be extended to other nonequilibrium critical
systems besides diffusive models ($Z=2$) and models above the upper 
critical dimension.

\vskip 0.5cm

\noindent
{\bf Acknowledgements:}\\
I thank N. Menyh\'ard, H. Hinrichsen and M. Henkel for the useful discussion
and A. Gambassi for the useful comments.
Support from Hungarian research funds OTKA (Grant No. T046129) is acknowledged.
The author thanks the access to the NIIFI Cluster-GRID, LCG-GRID 
and to the Supercomputer Center of Hungary.

\end{document}